\documentclass[11pt]{article}
\usepackage{kpfonts}
\usepackage{comment}
\usepackage{bm}
\usepackage{amssymb}
\usepackage{amsmath}
\usepackage{xfrac}
\usepackage{stmaryrd}
\usepackage{float}
\usepackage{etoolbox}
\usepackage{esvect}
\usepackage{amsthm}
\usepackage{bm}

\usepackage{graphicx}

\newtheorem{thm}{Theorem}

\newtheorem{lem}{Lemma}

\usepackage{enumerate}
\usepackage{paralist}
\usepackage{indentfirst}
\usepackage{setspace}	
\usepackage{natbib}	
\usepackage{bibentry}

\usepackage{appendix}
\usepackage{xspace}

\usepackage{color}
\usepackage[usenames, dvipsnames, svgnames, table]{xcolor} 

\usepackage{float}
\floatstyle{boxed}
\newfloat{aside}{h}{aside}
\usepackage{caption}
\usepackage{geometry}
\usepackage{algorithm}
\usepackage[noend]{algpseudocode}

\usepackage[pagebackref=true]{hyperref}
\hypersetup{
	pdfborder={0 0 0.5 [2 1]},
	colorlinks=true,
	linkcolor=blue,
	citecolor=Maroon,
	urlcolor=blue,
	linkbordercolor=blue,
	citebordercolor=PineGreen,
	urlbordercolor=blue
}

\usepackage[capitalize,nameinlink]{cleveref}
\crefname{ineq}{Inequality}{Inequalities}
\creflabelformat{ineq}{#2{\upshape(#1)}#3}
\Crefname{cor}{Corollary}{Corollaries}
\Crefname{conj}{Conjecture}{Conjectures}
\Crefname{thm}{Theorem}{Theorems}
\Crefname{rmk}{Remark}{Remarks}
\Crefname{prop}{Proposition}{Propositions}
\Crefname{lem}{Lemma}{Lemmata}
\Crefname{ex}{Example}{Examples}

\usepackage{tikz}

\newcommand{\pisd}{\mathrel u_{i}^{sd}}
\newcommand{\risd}{\mathrel {\underline u}_{i}^{sd}}

\DeclareMathOperator*\supp{supp}
\newcommand{\nsupp}[1]{|\text{supp}(#1)|}

\usepackage{amsmath}
\usepackage{accents}
\newlength{\dhatheight}

\newlength{\dbarheight}

\usepackage{soul} 
\usepackage{natbib}

\newcommand\continuous{normalized cone continuous\xspace}
\newcommand\continuity{normalized cone continuity\xspace}
\newcommand\Continuity{Normalized cone continuity\xspace}

\makeatletter
\long\def\@makecaption#1#2{%
  \vskip\abovecaptionskip
  \sbox\@tempboxa{\textbf{#1}: #2}%
  \ifdim \wd\@tempboxa >\hsize
    \textbf{#1:} #2\par
  \else
    \global \@minipagefalse
    \hb@xt@\hsize{\box\@tempboxa\hfil}%
  \fi
  \vskip\belowcaptionskip}
\makeatother

\title{
Ordinality in Random Allocation
}

\author{\begin{tabular}{ccc}
\begin{tabular}{c}
Eun Jeong Heo\\
{\footnotesize University of Seoul }\\
{\footnotesize eunjheo@uos.ac.kr }
\end{tabular}&$\qquad$&
\begin{tabular}{c}
Vikram Manjunath\\
{\footnotesize vikram@dosamobile.com}
\end{tabular}\\ & \\
          \end{tabular}
        }
\begin{document}
\sloppy
\maketitle

\begin{abstract}
In allocating objects via lotteries, it is common to consider
\emph{ordinal} rules that rely solely on how agents rank degenerate
lotteries. While ordinality is often imposed due to cognitive or
informational constraints, we provide another justification from an
axiomatic perspective: for three-agent problems, the combination of
efficiency, strategy-proofness, non-bossiness, and a weak form of continuity collectively implies ordinality.\\\bigskip

\noindent
{\it JEL} classification: C70, D61, D63.

\end{abstract}

\begin{center}
\end{center}

\section{Introduction}
\label{sec:introduction}

In this paper, we consider the \emph{random allocation} problem, where
each agent receives one object through a \emph{lottery}. A \emph{rule}
takes the agents' preferences over lotteries as input and selects an
allocation subject to feasibility constraints. This problem was
initially formulated and studied by
\cite{HyllandZeckhauserJPE1979}. Assuming agents report their Bernoulli utilities, they proposed a pseudomarket procedure. While this procedure achieves efficiency, it fails to guarantee truthful reporting of preferences unless there are sufficiently many agents. This occurs because the pseudomarket relies on cardinal information about preferences, making it susceptible to manipulation by individual agents.

Another line of research, initiated by
\cite{BogomolnaiaMoulinJET2001}, proposes using \emph{ordinal}
allocation rules, which rely solely on agents' preferences over the
objects (or equivalently, degenerate lotteries) as input.\footnote{The
  restriction to ordinal rules also appears in
  \cite{GibbardEconometrica1977} in the voting context.}  The focus on
ordinal rules is justified by the argument that agents may not have
expected utility preferences or may struggle to formulate and convey complete preferences over lotteries (\cite{KagelRoth1995}, \cite{BogomolnaiaMoulinJET2001}).

While many papers on the subject adopt this approach, there has not been sufficient discussion of why we should impose ordinality on the allocation rules from an axiomatic perspective. One might conjecture that an incentive requirement forces a rule to disregard the cardinal intensities of preferences, as implied by \cite{HyllandZeckhauserJPE1979}. For deterministic allocations, it is indeed clear that any strategy-proof rule should ignore cardinal information, but the situation is less straightforward when allocations are lotteries.

A recent paper by  \cite{EhlersMajumdarMishraSenJME2020} is
relevant. They investigate a similar question in a probabilistic
model. Since their 
results are for a single agent 
problem, they 
apply to a wide range of problems. They introduce a strong property
called ``uniform cone 
continuity'' and demonstrate that along with strategy-proofness this
property implies  ordinality. This form of continuity requires a rule to be uniformly continuous over the set of all utilities consistent with each ordinal preference. Without normalizing utilities, one can approach the utility vector $\vec{0}$ arbitrarily closely within the set of all utilities consistent with an ordinal preference. Thus, for sufficiently scaled-down utilities, a uniformly continuous rule must remain constant within this set. Strategy-proofness then ensures that scaling the utilities back has no effect on the rule's choice.

Turning to the random allocation problem, we introduce a mild and
natural continuity requirement over the domain of normalized Bernoulli utilities
without ties.
The space of all profiles of Bernoulli utilities, when we have $n$
agents and $m$ objects, is $\mathbb R^{mn}$. By the Birkhoff-von
Neumrann Theorem, we can represent the set of all random allocations
by the set of bistochastic matrices
\citep{Birkhoff:1946,vonNeumann:1953}, a subset of $\mathbb
R^{mn}$. Since a rule is a mapping from utility profiles to random
allocations, one could  consider the standard notion of continuity of
a function from $\mathbb R^{mn}$ to itself.
However, any efficient rule violates this notion of
continuity. Moreover, certain discontinuities at the boundaries
between Bernoulli utilities that rationalize one strict order and
those that 
rationalize another strict order are unavoidable.\footnote{Take for instance
  a situation where two agents rank a pair of objects in the opposite
  wqy---efficiency says each of them gets their preferred object. So,
  along any sequence of such preference profiles, efficiency requires
  the rule to be constant. 
  Efficiency says nothing when they are both indifferent, so a rule
  can pick anything. Along different sequences, both tending towards
  indifference in the limit, the rule could be constant at different
  deterministic allocations. So, discontinuity is unavoidable at the
  indifferent profile. Formally,  consider a
  pair of agents $i$ and $j$, a pair of  objects $a$ and $b$, and a
  pair of sequences $\{(u_i^k, u_j^k)\}_{k=1}^\infty$ and  $\{(v_i^k,
  v_j^k)\}_{k=1}^\infty$ 
  such that $u_i^k(a) = v_j^k(a) = u_j^k(b) = v_i^k(b) =\frac{1}{k}$
  and $u_i^k(b) = v_j^k(b) = u_j^k(a) = v_i^k(a) =0$. Efficiency
  says that for each 
  $k < \infty$, at the profile $(u_i^k,u_j^k)$ the rule allocates $a$
  to $i$ with certainty and $b$ to $j$ with certainty. Similarly at
  $(v_i^k, v_j^k)$ it allocates $b$ to $i$ and $a$ to $j$. Yet, these
  sequences converge to the same limit, so the rule cannot be
  continuous at  the profile where both agents assign the same utility
  to each object.}
We therefore exclude these  boundaries from the domain. Continuity
over this domain is weaker than continuity over the full
domain. Moreover, we only require continuity with respect to a single
agent's utilities, not the profile as a whole.

For three-agent problems,  we show that this milder continuity
requirement and strategy-proofness, along with efficiency and
non-bossiness, jointly imply ordinality. Whether this generalizes
to more than three agents remains an open question. Nonetheless, this
finding provides a foundation for ordinality in random allocations. 

As an extension, we show that even on larger domains than expected
utility preferences, rules satisfying our axioms necessarily ignore
non-ordinal information.

The remainder of the paper is organised as follows: In
\cref{sec:model-1} we formally define the model where agents compare
lotteries based on expected utility. We define our axioms
in \cref{sec:eff-axiom}. We prove our main result on ordinality in
\cref{sec:main-result}. Finally, in \cref{sec:nonEU} we show that this
result extends to larger domains of preferences. 

\section{The Model}
\label{sec:model-1}
Let $A$ be a  set of at least three  objects and
$N$ be a set of at least three  agents. There is one unit of supply of
each object and
there are as many objects as there are agents, so $|A| = |N| =
3$.\footnote{When there are fewer than three objects, ordinality does
  not imply any loss of generality for this model. So, we do not
  consider the case of two agents and two objects.}
Let $\Delta(A)$ be the set of lotteries over $A$. As it does not cause
confusion, we identify
degenerate lotteries in $\Delta(A)$ with $A$ itself. So, for each
$a\in A$, we denote the lottery that places probability one on $a$ by
$a$.
Let $\mathcal U$ be the domain of all possible Bernoulli utilities with no
ties over degenerate lotteries---that is, for each $i\in N$, each
$u_i\in \mathcal U$, 
and each pair $a,b\in A$, $u_i(a) \neq u_i(b)$. For an  agent $i\in N$
whose 
Bernoulli utility is $u_i\in \mathcal U$,  her expected utility from
$\pi_i\in \Delta(A)$ is  $u_i(\pi_i) = \sum_{a\in A}u_i(a)\pi_{ia}$.

Let $\mathcal P$ be the set of strict
preference relations (complete, transitive, and antisymmetric binary
relations) over $A$.
We say that $u\in \mathcal U$ is \textbf{consistent with}
$P\in\mathcal P$ if for each pair $a,b\in A$, $u(a) > u(b)$ if
and only if $a \mathrel P b$.
For each $P\in \mathcal P$, let
{\boldmath $\mathcal U_{P}$} be the set of utilities that are
consistent with $P$.
Our assumption of no ties on degenerate lotteries ensures
that $(\mathcal U_{P})_{P\in  \mathcal P}$ is a
partition of $\mathcal U$.
Define, for each $P\in\mathcal P^N$, $\mathcal U_P^N = \times_{i\in N}
\mathcal U_{P_i}$ as the set of utility profiles consistent with the
preference profile $P$. Again, $(\mathcal U_{P}^N)_{P\in \mathcal P^N}$ is a
partition of $\mathcal U^N$.

An allocation $\pi$ specifies, for each agent $i\in N$ and each object
$a\in A$, the probability that $i$ receives $a$: $\pi_{ia}$. Each
agent  consumes  one object and each object is
consumed by at most one agent. So,  for each $i\in N$, $\sum_{a\in A}
\pi_{ia} = 1$ and for each $a\in A$, $\sum_{i\in N}\pi_{ia}=
1$. Thus, the  set of all allocations, $\Pi$, is the set of all
bistochastic matrices.\footnote{Every $\pi\in \Pi$ can be expressed as
a lottery over  permutation matrices
\citep{Birkhoff:1946,vonNeumann:1953}. So, one can  always
realize the lotteries of all agents jointly in a way that is
consistent with their individual lotteries.}

A rule, $\varphi: \mathcal U^N \to \Pi$, associates every profile of
utilities with an allocation. For each $u \in \mathcal U^N$, each
$i\in N$, and each $a\in A$,  we denote by $\varphi_i(u)$ the
lottery that  $i$ receives and by $\varphi_{ia}(u)$ the
probability that $i$ receives $a$.

\subsection{Axioms}
\label{sec:eff-axiom}

Consider an allocation $\pi\in \Pi$ and $u\in \mathcal U^N$. We would
not consider $\pi$ an
effective use of the available resources if there were an opportunity
to make some agent $i\in N$ better off, via $\pi'\in \Pi$,  where
witching from $\pi$ to $\pi'$ does not decrease the expected utility
of any  other agent $j\in N\setminus \{i\}$.\footnote{See
  \cite{HeoManjunathAlva2025} for 
  a discussion of the lineage of this notion of efficiency from Pareto
  and Edgeworth to  its formulation in the literature on random
  allocation.} We say that
$\pi'$ \textbf{dominates} $\pi$ at $u$ if there is an agent $i\in N$
such that $u_i(\pi_i')> u_i(\pi_i)$ and
for every other agent $j\in N\setminus\{i\}$, $u_j(\pi_j')\geq
 u_j(\pi_j) $.
An allocation $\pi$ is \textbf{efficient} at $u$ if it is not dominated
by  any other allocation.
A rule $\varphi$ is efficient if, for
each $u\in \mathcal U^N$, $\varphi(u)$ is efficient at $u$.

A rule $\varphi$  is \textbf{strategy-proof} if it is a weakly
dominant strategy for each agent to submit their true utility function
in the direct revelation game associated with it.  That is, for each
$u\in \mathcal U^N$, each $i\in N$, and each $u_i'\in \mathcal U$,
\[
  u_i(\varphi_i(u_i, u_{-i})) \geq u_i(\varphi_i(u_i', u_{-i})).\footnotemark
\]\footnotetext{Since we have discussed the origins of efficiency, for
  the sake of consistency, we do the same for  the other axioms
  as well. This formulation of immunity to manipulation
  originates in \cite{BlackJPE1948} and \cite{FarquharsonOEP1956}.}

Next, we describe a   regularity property. A rule
$\varphi$  is \textbf{non-bossy} if for each $u\in \mathcal
  U^N$, each $i\in N$, and each $u_i'\in \mathcal U$ such that
  $\varphi_i(u) = \varphi_i(u_i',u_{-i})$, we have $\varphi(u) =
  \varphi(u_i',u_{-i})$.\footnote{Non-bossiness was first defined in
    \cite{SatterthwaiteSonnenscheinRES1981}.}

  Finally, we consider a mild form of continuity.
Before defining it, we observe that 
$\mathcal U$ is not a compact set as it is missing the
boundaries of $ {\mathcal U}_{P_i}$ for each $P_i\in\mathcal
P$. This is true even  if we consider the normalized set of Bernoulli
utilities in it that sum to one, $\overline
{\mathcal U} = \{u_i\in \mathcal U: \sum_{a\in A} u_i(a) = 1\}$.
We say that a rule $\varphi$ is \textbf{\continuous} if, for 
each agent $i\in N$, each profile of others' preferences $u_{-i}\in
\mathcal U^N$, and each sequence of utility
functions for $i$, $\{u_i^k\}_{k=1}^\infty\in \overline{\mathcal U}^N$
with  limit $u_i\in\overline{\mathcal U}$, $\varphi(u_i, u_{-i}) =
\lim_{k\to\infty}\varphi(u_i^k, u_{-i})$.\footnote{To the best of our knowledge,
  continuity as a property of for social choice rules dates back to
  \cite{ChichilniskyAM1980}.} Since Bernoulli
  utilities are elements of Euclidean space, convergence is with
  respect to Euclidean distance. Terminologically, by
 ``normalized'' we mean that our continuity requirement only applies
 over the normalized utilities in $\overline {\mathcal U}$ as opposed
 to all of  $\mathcal U$. The reference to ``cone''  is
 because we are only concerned with changes to an agent's utility that
 remain within the cone of utilities that are consistent with the same
 ordinal preference.
  Since we only consider sequences  in
$\overline{\mathcal U}$ and do not demand uniform continuity, this is
considerably weaker than the  
\emph{uniform cone continuity} of
\cite{EhlersMajumdarMishraSenJME2020}. 

A normative interpretation of
\continuity is that small errors in 
stating   one's Bernoulli utilities ought not have  large  effects on
the  final allocation.

\section{Main Result}\label{sec:main-result}
A rule is \textbf{ordinal} if it  depends only on how the input
utilities rank degenerate lotteries. Formally, for each pair of profiles $u, u'\in \mathcal U^N$,
if there is $P\in\mathcal P^N$ such that $u, u'\in
\mathcal U^N_{P}$, then $\varphi(u) = \varphi(u')$. Thinking of this
as an informational constraint, it is equivalent to saying that a rule
is measurable with respect to the $\sigma$-algebra generated by
$\{\mathcal U_{P}^N\}_{P\in \mathcal  P^N}$.
 We show that the   axioms that we listed in the previous section
 imply that a rule is ordinal.

\begin{thm}\label{ordinality theorem} Suppose $|A| =
  |N| = 3$.
  Every \continuous, efficient, strategy-proof,
  and non-bossy rule is ordinal.
\end{thm}

This result highlights that ordinality is derived from our relatively
weak continuity property and three standard axioms in random
allocations. The proof for three agents and three objects is rather
involved and extension to more than three agents does not seem
tractable. We are therefore  unable to generalize it to arbitrary finite economies. We leave this as an open question.

We prove a series of lemmata to establish \Cref{ordinality theorem}.
Here  is some notation we use in these proofs:
\begin{enumerate}
\item  For each $u\in \mathcal U^N$ and each $i\in N$,  $ \bm{P(u_i)} \in
  \mathcal   P$ is such that $u_i\in\overline{\mathcal U}_{P(u_i)}$.
\item For each $u\in \mathcal U^N$, $\bm{P(u)} = (P(u_i))_{i\in N}$.
\item   For each $\pi\in \Pi$ and $i\in N, \bm{\supp(\pi_i)} = \{a\in A:
  \pi_{ia} > 0\}$ is the support of $\pi_i$.
\item Given a pair  $a,b\in A$, $\bm{[a,b]} = \{\pi\in \Delta(A):
  \pi_{a} + \pi_{b} = 1\}$.

\item Given a pair  $a,b\in A$, $\bm{(a,b]} = [b,a) = \{\pi\in [a,b]: \pi_a > 0\}$.

\item Given a pair  $a,b\in A$, $\bm{(a,b)} = \{\pi\in (a,b]: \pi_b > 0\}$.
\item Given $u_i\in \mathcal U$, such that $u_i(a) > u_i(b) > u_i(c)$,
 the \textbf{rate of middle substitution} of $u$ is
  \[
    \bm{\mu(u_i)} = \frac{u_{ib} - u_{ic}}{u_{ia} - u_{ic}}.\footnotemark
  \] \footnotetext{Given a Bernoulli  utility
$u_i\in\mathcal U$ for $i$ where $a$ is the best object and $c$ is the
worst, $\frac{u_{ib} - u_{ic}}{u_{ia} - u_{ic}}$ is $i$'s rate of
substitution of
$a$ for $b$ where we adjust the amount of $c$ so that the sum of
probabilities is 1.}
\item A pair $u_i, u_i'\in\mathcal U$ are \textbf{effectively the
    same} if $P(u_i)
  = P(u_i')$ and $\mu(u_i) = \mu(u_i')$. Denote this by $u_i \bm{\sim}
  u_i'$. The normalization that we use to define
  $\overline{\mathcal  U}$ in \cref{sec:model-1} ensures that there are no two
  distinct utilities in  $\overline{\mathcal U}$ that are effectively
  the same.
\end{enumerate}

\Continuity is defined only for utility profiles in $\mathcal U^N$,
that too, only with respect to  sequences of individual utilities in $\overline
{\mathcal U}$. Our first task is to show that a strategy-proof,
non-bossy, and \continuous rule only responds to how each agent's
utility function orders lotteries. So, if we replace an agent $i$'s
utility function $u_i$ with another utility function $u_i'$ that is
effectively the same as $u_i$, this does not affect what the rule
chooses.

\begin{lem}
  \label{lem: lottery-ordinal utility}
  Suppose $\varphi$ is strategy-proof, non-bossy, and
  \continuous. Let $i\in N$,   $u\in \mathcal U^N$, and $\overline u_i\in
  \overline{\mathcal U}$. If  $u_i\sim \overline u_i$ then $\varphi(u_i, u_{-i}) =
  \varphi(\overline u_i, u_{-i})$.
\end{lem}
\begin{proof}
Let $\pi =  \varphi(u_i, u_{-i})$ and $\overline \pi=\varphi(\overline
u_i, u_{-i})$. Suppose $\pi\neq \overline \pi$. Since $\varphi$ is
non-bossy, $\pi_i \neq \overline \pi_i$. Label the objects so that
$u_i(a) > u_i(b) > u_i(c)$.  Since $\varphi$ is strategy-proof,
$u_i(\pi_i) = u_i(\overline \pi_i)$ (equivalently, $\overline
u_i(\pi_i) = \overline u_i(\overline \pi_i)$). Since  $u_i$ is linear
and $\pi_i,\overline\pi_i\in \Delta(A)$, $\pi_{ib}
\neq \overline \pi_{ib}$. Suppose $\pi_{ib} > \overline \pi_{ib}$ (the
proof is analogous if $\pi_{ib} < \overline \pi_{ib}$).

Let $\alpha\in (0,1)$ be such that $u_i \sim \overline u_i \sim (1,\alpha,
0)$.\footnote{We represent a  Bernoulli utility $u_i$ as a  vector
  $(u_{i}(a), u_{i}(b), u_{i}(c))$.}

For some $\alpha' \in (0, \alpha)$, let $\overline u_i' \in \overline
{\mathcal U}$ be such that $\overline u_i' \sim
(1,\alpha',0)$ and let $\overline \pi' = \varphi(\overline
u_i',u_{-i})$. Since $\varphi$ is strategy-proof, $(1,\alpha, 0)
\cdot (\overline \pi_i - \overline \pi_i') \geq 0$ and $(1,\alpha', 0)
\cdot (\overline \pi_i' - \overline \pi_i) \geq 0$. Thus, $(\alpha -
\alpha')(\overline \pi_{ib} - \overline \pi_{ib}') \geq 0$. Since
$\alpha > \alpha'$, this means $\overline \pi_{ib} \geq \overline \pi_{ib}'$.

For some $\alpha''\in (\alpha,1)$, let
$\overline u_i'' \in \overline {\mathcal U}$ be such that  $\overline
u_i'' \sim (1,\alpha'',0)$ and $\overline\pi'' = \varphi(\overline
u_i'',u_{-i})$. By the same argument as above, we conclude
that $\pi_{ib} \leq \overline \pi_{ib}''$. In particular, this implies
that for each $\tilde u_i \in \overline {\mathcal U}$ with $\tilde
u_i(a) > \tilde u_i(b) > \tilde u_i(c)$, $\varphi_{ib}(\tilde
u_i,u_{-i})\notin (\overline \pi_{ib}, \pi_{ib})$.

Finally, for each $\beta\in [0,1]$, let
\[
  \overline u_i^\beta = \beta \overline u_i' + (1-\beta) \overline
  u_i'' \in \overline {\mathcal U} \hspace{0.5in}\text{ and }\hspace{0.5in}\overline \pi^\beta =
  \varphi(\overline u_i^\beta, u_{-i}).
\]
When $\beta = 0$, $\overline \pi_{ib}^\beta =
\overline \pi_{ib} \leq \overline \pi_{ib}'$ and when $\beta = 1$,
$\overline \pi_{ib}^\beta = \overline \pi_{ib}'' \geq \pi_{ib}$.
Since there is no $\beta$ such that $\pi_{ib}^\beta \in  (\overline
\pi_{ib}, \pi_{ib})$, this contradicts the \continuity of $\varphi$.
\end{proof}

\cref{lem: lottery-ordinal utility} tells us that for  a  strategy-proof,
non-bossy, and \continuous rule, we can switch between any pair of
utility functions or profiles that are effectively the
same without affecting the choice of the rule.

Increasing the utility from a specific object that an agent receives
with positive probability is a sort of monotonic transformation of
that agent's preferences over lotteries. However, due to the linearity
of indifference curves when utilities are in $\mathcal U$, upper
contour sets are not necessarily nested
when such a change occurs. Nonetheless, if an agent consumes a
positive amount of only two objects that are contiguous in their
ordinal ranking, we do see such a nesting. Our next lemma says that
such changes do not affect an agent's allocation.

\begin{lem}\label{lem: maskin invariance on boundary} Suppose {$\varphi$ is strategy-proof.}
Let   {$u\in \mathcal U^N$ be such that for $i\in N$, $u_{i}(a) > u_{i}(b) >
    u_{i}(c)$,  $u_i'\in \mathcal U$ be such that $P(u_i') = P(u_i)$
    and
    $\mu(u_i') > \mu(u_i)$.
}
  {  If $\varphi_i(u)  \in [a,b]\cup [b,c]$ then $\varphi_i(u_i',u_{-i})
    = \varphi_i(u)$}.
\end{lem}
\begin{proof}
  Since $P(u_i') = P(u_i)$ and    $\mu(u_i') > \mu(u_i)$, by
  \cref{lem: lottery-ordinal utility} it is
  without loss of generality to suppose that there is   $\alpha\in
  (0,u_{i}(a) - u_{i}(b))$ such that
  $u_{i}'(a) = u_{i}(A), u_{i}'(b) = u_{i}(b) + \alpha,$ and
  $u_{i}'(c)=u_{i}(c) $.

  Let $\pi_i = \varphi_i(u) $ and $\pi_i' = \varphi_i(u_i',u_{-i})$.
  By strategy-proofness,
  \begin{equation*}
    \pi_{ia} u_{i}(a) + \pi_{ib} (u_{i}(b) + \alpha) + \pi_{ic} u_{i}(c) \leq \pi_{ia}'u_{i}(a) +\pi_{ib}' (u_{i}(b)+ \alpha) + \pi_{ic}' u_{i}(c).
  \end{equation*} Otherwise, $i$ would profitably report  $u_i$ at
  $(u_i',u_{-i})$. We re-write this as follows:
  \begin{equation}
    \label[ineq]{ineq: Maskin 1}
    u_{i}(a) (\pi_{ia} - \pi_{ia}') + u_{i}(b)(\pi_{ib} - \pi_{ib}') +
    u_{i}(c)(\pi_{ic} - \pi_{ic}') \leq  \alpha(\pi_{ib}' - \pi_{ib}).
  \end{equation}
Similarly, strategy-proofness implies  that
  \begin{equation*}
    \pi_{ia} u_{i}(a) + \pi_{ib} u_{i}(b) + \pi_{ic} u_{i}(c) \geq \pi_{ia}'u_{i}(a) +\pi_{ib}' u_{i}(b) + \pi_{ic}' u_{i}(c).
  \end{equation*}
  Equivalently,
  \begin{equation}
    \label[ineq]{ineq: Maskin 2}
    u_{i}(a) (\pi_{ia} - \pi_{ia}') + u_{i}(a)(\pi_{ib} - \pi_{ib}') +
    u_{i}(c)(\pi_{ic} - \pi_{ic}') \geq 0.
  \end{equation}
  From \Cref{ineq: Maskin 1,ineq: Maskin 2}, $\alpha(\pi_{ib}' -
  \pi_{ib})   \geq 0$. Since $\alpha > 0$, this means $\pi_{ib}' \geq
  \pi_{ib}$.

  If $\pi_i \in [a,b]$, then $\pi_{ib}' \geq \pi_{ib}$ implies that
  $\pi_{ia}' \leq \pi_{ia}$. Consider $\tilde \pi_i$ such that
  $\tilde\pi_{ib} = \pi_{ib}'$ and $\tilde \pi_{ia} = 1-\tilde
  \pi_{ib}$. By assumption on $u_i',$ $u_i'(\tilde \pi_i) \geq
  u_i'(\pi_i').$ However, since $\pi_i\in [a,b]$, if $\pi_{ib}'\neq
  \pi_{ib}$,  $u_i'(\pi_i) > u_i'( \tilde\pi_i)$, violating
  strategy-proofness   of $\varphi$. Thus, $\pi_{ib}'=
  \tilde \pi_{ib} = \pi_{ib}$. Given this, if $\pi_{ia}' < \pi_{ia}$,
  then   $u_i'(\pi_i') < u_i'( \pi_i)$, again violating
  strategy-proofness of $\varphi$. Thus, $\pi_i = \pi_i'$.

  If $\pi_i \in [b,c]$, then $\pi_{ib}' \geq \pi_{ib}$ implies that
  $\pi_{ic}' \leq \pi_{ic}$. Consider $\tilde \pi_i$ such that
  $\tilde\pi_{ib} = \pi_{ib}'$ and $\tilde \pi_{ic} = 1-\tilde
  \pi_{ib}$. By assumption on $u_i,$ $u_i(\tilde \pi_i) \leq
  u_i(\pi_i').$ However, since $\pi_i\in [b,c]$, if $\pi_{ib}'\neq
  \pi_{bi}$,  $u_i(\tilde\pi_i) > u_i(\pi_i)$ meaning that
  $u_i(\pi_i') >   u_i(\pi_i)$, violating strategy-proofness  of
  $\varphi$.  Thus, $\pi_{ib}'=
  \tilde \pi_{ib} = \pi_{ib}$. Given this, if $\pi_{ic}' < \pi_{ic}$,
  then   $u_i(\pi_i' )> u_i(\pi_i)$, again violating
  strategy-proofness of $\varphi$. Thus, $\pi_i = \pi_i'$.
\end{proof}

Next, we show that if two agents have the effectively the same
utility function, then
each agent receives an allocation on a specific boundary of the
simplex as long as the third does not receive their middle object in
entirety.

\begin{lem}
  \label{lem: identical utilities for two agents}
  Suppose {$\varphi$ is
    {efficient} and  {\continuous}.
  } Let
  {$u\in \mathcal U^N$ be such that for a distinct pair $i,j\in N$,
    $u_i \sim u_j $, $u_{ia} >
    u_{ib} > u_{ic}$, and  $\pi = \varphi(u)$.}
  {
    If $\pi_{ib} + \pi_{jb} >0$, then $\pi_i, \pi_j\in [a,b]\cup[b,c]$.
  }
\end{lem}
\begin{proof}
Suppose, to
the contrary, that $ \pi_{ia}> 0$ and $\pi_{ic} > 0$.
If $\pi_{jb} = 0$, then by feasibility, $\pi_{ja} >0$ and $\pi_{jc} >
0$. In this case, since $\pi_{ib} + \pi_{jb} > 0$,  $\pi_{ib} >0$.
Since we can switch the names of $i$ and $j$ in such a case, it is
 without loss of generality to suppose, for the sake of
contradiction, that $\pi_{ia} > 0, \pi_{ic} > 0,$ and $\pi_{jb} > 0$.

By \cref{lem: lottery-ordinal utility} we may assume that $u_i\in
\overline{\mathcal U}$.    Let $\delta \in (0,  u_{ia}-  u_{ib})$ and let
 be such that $ u_{i}' =\frac{1}{1+\delta}
  ( u_{i}(a),   u_{i}(b) + \delta, u_{i}(c)) \in\overline {\mathcal U}$.
  Let $ \pi' = \varphi( u_i',  u_{-i})$.
  By \continuity of $\varphi$, since
  $\pi_{ia} > 0, \pi_{ic} > 0,$ and $\pi_{jb} > 0$, if $\delta$ is
sufficiently small,   $\pi_{ia}' > 0, \pi_{ic}' > 0,$ and $\pi_{jb}' > 0$.

Now consider an arbitrarily small
$\varepsilon>0$. It is feasible to transfer $\varepsilon$ share of $b$
from $j$ to $i$ in exchange for $\varepsilon \mu(u_i)$ share of
$a$ and $\varepsilon\left(1-\mu(u_i)\right)$ share of
$c$. Such a trade between $i$ and $j$ would leave $j$
indifferent and make $i$ better off, contradicting the efficiency of
$\varphi$.
Thus,
$i\in N$, $\pi_i \in [a,b]\cup[b,c]$.
\end{proof}

Several of the lemmata below are like the next one. They show that the
rule must choose the same allocation at certain ordinally equivalent
utility profiles.  The first of such cases is where an agent who
receives either their most or least preferred object with certainty
changes their Bernoulli utilities without affecting their ordinal
ranking of the object.

\begin{lem}
  \label{lem: top or bottom}
Suppose {$\varphi$ is
    {strategy-proof}.
  } Let
  {$u\in \mathcal U^N$ be such that for $i\in N$,
    $u_{i}(a) >    u_{i}(b) > u_{i}(c)$ and $u_i'\in \mathcal U$ be such
    that $P(u_i') = P(u_i)$.}
  {
    If $\varphi_{ia}(u)= 1$ or $\varphi_{ic}(u) = 1$, then
    $\varphi_i(u_i', u_{-i}) = \varphi_i(u)$.
  }
\end{lem}
\begin{proof}
    If $\varphi_{ia}(u)= 1$ but $\varphi_{ia}(u_i', u_{-i}) < 1$, then
    since $u_{i}'(a) > u_{i}'(\pi_i)$ for each $\pi_i\in
    \Delta(A)\setminus \{a\}$,
    $u_i'(\varphi_i(u)) > u_i'(\varphi_i(u_i',u_{-i}))$,
    violating strategy-proofness of $\varphi$.

    If $\varphi_{ic}(u)= 1$ but $\varphi_{ic}(u_i', u_{-i}) < 1$, then
    since $u_{i}'(c) < u_{i}'(\pi_i)$ for each $\pi_i\in\Delta( A)\setminus \{c\}$,
    $u_i(\varphi_i(u)) < u_i(\varphi_i(u_i',u_{-i}))$,
    violating strategy-proofness of $\varphi$.
\end{proof}

The next step is to extend  \cref{lem: identical utilities for two
  agents} from the case where two agents have effectively the same
utility to where they have the same ordinal ranking.

\begin{lem}
  \label{lem: positive b implies ab u bc}
Suppose {$\varphi$ is     {efficient}, {strategy-proof},
    {non-bossy},
and {\continuous}.
  } Let
  {$u\in \mathcal U^N$ be such that for a distinct pair $i,j\in N$,
    $P(u_i) = P( u_j) $, $u_{i}(a) >
    u_{i}(b) > u_{i}(c)$, and
    $\pi = \varphi(u)$.}
  {
    If $\pi_{ib} + \pi_{jb} >0$, then $\pi_i, \pi_j\in [a,b]\cup[b,c]$.
  }
\end{lem}
\begin{proof}
  Without loss of generality, suppose that $\mu(u_j) < \mu(u_i)$. We
  first consider the case that $\pi_{jb} > 0$.
  By \cref{lem: lottery-ordinal utility} we may suppose, without loss
  of generality, that $u_i,  u_j\in \overline {\mathcal   U}$.
  Efficiency dictates
  that $\pi_{ia} = 0$ or $\pi_{ic} = 0$ for otherwise, transferring a
  small $\varepsilon$ amount  of $b$ from $j$ to $i$ in return for
  $\varepsilon\mu(u_j)$ $a$ and $\varepsilon(1-\mu(u_j))$ of $c$ would
  yield a Pareto-improvement. Thus, $\pi_i\in [a,b]\cup[b,c]$.
  If $\pi_{ia} = 1$ or   $\pi_{ic}  = 1$, by \cref{lem: top or
    bottom}, setting $u_i' = u_j$,
  $\varphi(u_i', u_{-i}) = \pi$, so we are done by \cref{lem:
    identical utilities for two agents}. That leaves us with the case
  of   $\pi_i\in  (a,b) \cup (b,c)$.  For each $\alpha\in[0,1]$, let
  \[
    u_i^\alpha = \alpha u_j + (1-\alpha)u_i\in \overline{\mathcal
      U}\hspace{0.5in}\text{ and
    }\hspace{0.5in}\pi^\alpha =
    \varphi(u_i^\alpha, u_{-i}).
  \]
  For each $\alpha \in [0,1)$, since $\mu(u_j) < \mu(u^\alpha_i)$, as
  explained above, $\pi^\alpha_{jb}>0$ implies that $\pi_i^\alpha\in
  (a,b)\cup(b,c)$. If $\pi^1_{jb} = 0$, then by \continuity of
  $\varphi$, since $\pi^0_{jb} > 0$, there is some $\alpha^* <1$ such
  that $\pi^{\alpha^*}_j  \neq \pi_j^0 = \pi_j$ but
  $\pi^{\alpha^*}_{jb} > 0$.  As argued above, this implies
  $\pi^{\alpha^*}_i \in   (a,b)\cup (b,c)$.  If $\pi_i^{\alpha^*} \neq
  \pi_i^0$, then because $\pi_i^{\alpha^*}, \pi_i^0\in
  (a,b)\cup(b,c)$, either $u_i^{\alpha^*} ( \pi_i^{\alpha^*}) <
  u_i^{\alpha^*}( \pi_i^{0})$ or $u_i^{0} ( \pi_i^{\alpha^*}) >
  u_i^0( \pi_i^{0})$, contradicting the strategy-proofness of
  $\varphi$. Thus, $\pi_i^1 = \pi_i^0$, so by non-bossiness,
  $\pi^1 = \pi^0$. By \cref{lem: identical utilities for two agents}
  we then have $\pi_i^1, \pi_j^1 \in
  [a,b]\cup[b,c]$. Thus, we have established that if $\pi_{jb} > 0$,
  then   $\pi_i, \pi_j \in [a,b]\cup[b,c]$.

  Now we consider the case where $\pi_{jb} = 0$, so that $\pi_{ib} >
  0$. Analogous to the argument above,  if $\pi_{ja} = 1$ or $\pi_{jc}
  = 1$, then \cref{lem: top or bottom} allows us to replace  $u_j$
  with $u_i$ and we  conclude by appealing to  appealing  to
  \cref{lem: identical utilities for two agents}. So, suppose that
  $\pi_j \in (a,c)$.
  For each $\alpha\in[0,1]$, let
  \[
    u_j^\alpha = \alpha u_i + (1-\alpha)u_j\in \overline {\mathcal U}\hspace{0.5in}\text{ and
    }\hspace{0.5in}\pi^\alpha =
    \varphi(u_j^\alpha, u_{-j}).
  \]
  If $\pi^1_{ib} + \pi^1_{jb} >0$, then by \cref{lem:
    identical utilities for two agents}, $\pi^1_j \in
  [a,b]\cup[b,c]$. However, since $\pi^0_j\in (a,c)$, this means for
  some $\alpha$, either $\pi^\alpha_{ja} = 1$ or  $\pi^\alpha_{jc} =
  1$, contradicting \cref{lem: top or bottom}. Thus, $\pi^1_{ib} =
  \pi^1_{jb} = 0$ and $\pi^1_j \in (a,c)$.
  If for each  $\alpha$, $\pi_{jb}^\alpha  = 0$, then $\pi_{jb}^\alpha
  \in [a,c]$ and strategy-proofness implies that $\pi_j^\alpha =
  \pi_j^0$. So, by non-bossiness $\pi^\alpha = \pi^0 = \pi$,
  contradicting $\pi^1_{ib} = 0$.
  Finally, if there is $\alpha^*$ such that $\pi^{\alpha^*}_{jb} > 0$,
  then   by the argument in the first paragraph of this proof,
  $\pi^{\alpha^*}_{j} \in [a,b]\cup[b,c]$. Since $\pi_j^0\in (a,c)$, this
  means there is some $\alpha\in(0,\alpha^*)$ such that
  $\pi_{ja}^\alpha = 1$ or   $\pi_{jc}^\alpha = 1$, and we reach a
  contradiction by appealing to  \cref{lem: top or bottom,lem:
    identical utilities for two agents} as above.
\end{proof}

If two agents have the same ordinal ranking and at least one of them
receives their middle ranked object with positive probability, then
the next two lemmata state that the rule must ignore the non-ordinal
changes to either of their utility
functions. 
\begin{lem}\label{lem: invariance for one
    agent} Suppose {$\varphi$ is     {efficient}, {strategy-proof},
    {non-bossy}, and {\continuous}. } Let
  {$u\in \mathcal U^N$ be such that for some pair $i,j\in N$, $P(u_i)
    = P(u_j)$ where $u_{i}(a) > u_{i}(b) > u_{i}(c)$, $\pi =
    \varphi(u)$,
    and $u_i'\in\mathcal U$
    be such that $P(u_i') = P(u_i)$.}
  {  If $\pi_{ib} + \pi_{jb} > 0$,  then
     $\varphi(u'_i, u_{-i}) = \varphi(u)$. Moreover,
      $\pi_i, \pi_j\in [a,b]\cup[b,c]$.}
\end{lem}
\begin{proof}
  If $\pi_{ia} = 1$ or $\pi_{ic} = 1$ then we are done by \cref{lem:
    top or bottom}. So, suppose that $\pi_i\in (a,b]\cup[b,c)$. By
  \cref{lem: lottery-ordinal utility} we may assume without loss of
  generality that $u_i, u_i'\in
  \overline {\mathcal U}$. So,
  for
  each $\alpha\in[0,1]$,  let
  \[
    u_i^\alpha = \alpha u_i' + (1-\alpha)u_i\in   \overline {\mathcal
      U}\hspace{0.5in}\text{ and
    }\hspace{0.5in}\pi^\alpha =
    \varphi(u_i^\alpha, u_{-i}).
  \]
If $\pi^1_{ib}>0$ then by \cref{lem: positive b implies ab u bc},
$\pi_i1\in[a,b]\cup[b,c]$ and so by
  strategy-proofness, $\pi_i^1 = \pi_i$ and we are done.  So, suppose
  $\pi^1_{ib} = 0$ so that
  $\pi^1_i \in (a,c)$. In this case, \cref{lem: positive b implies ab
    u bc} implies that $\pi_{ib}^1 + \pi_{jb}^1 = 0$.
  Since, for each
  $\alpha$ such that $\pi_{ib}^\alpha + \pi_{jb}^\alpha > 0$,
  by  \cref{lem: positive b implies ab
    u bc}, $\pi^\alpha_i\in
  [a,b]\cup[b,c]$ but  $\pi^1_i \in (a,c)$, there exists, by \continuity
  of $\varphi$, some
  $\alpha^*$ such that $\pi^{\alpha^*}_{ia} = 1$ or
  $\pi^{\alpha^*}_{ic} = 1$. However, \cref{lem: top or bottom} then implies that
  $\pi_i^1 = \pi_i^{\alpha^*}$, contradicting $\pi_i^i \in (a,c)$.
\end{proof}

\begin{lem}\label{lem: same oder two agents}Suppose {$\varphi$ is
    {efficient}, {strategy-proof},
    {non-bossy}, and {\continuous}. } Let
  {$u\in \mathcal U^N$ be such that for some pair $i,j\in N$, $P(u_i)
    = P(u_j)$ where $u_{i}(a) > u_{i}(b) > u_{i}(c)$.}
  {  If $\varphi_{ib}(u) + \varphi_{jb}(u) > 0$, then
      for each $u'_i, u_j' \in \mathcal U$ such that $P(u_i') =
      P(u_j') = P(u_i)$, setting $u' = (u_i', u_j', u_{-\{i,j\}})$, $\varphi(u') = \varphi(u)$. Moreover,
      $\varphi_i(u), \varphi_j(u) \in [a,b]\cup[b,c]$.}
\end{lem}
\begin{proof}
  Let $\pi  = \varphi(u)$. By \cref{lem: positive b implies ab u bc},
  $\pi_i,\pi_j\in  [a,b]\cup[b,c]$. Since $\pi_{ib}+  \pi_{jb} > 0$,
  by \cref{lem: invariance for one  agent},
  $\varphi_i(u_i', u_{-i}) = \pi_i$ and by non-bossiness,
  $\varphi(u_i', u_{-i}) = \varphi(u)$. Repeating this argument for
  $j$, $\varphi(u') = \varphi(u_i', u_{-i}) = \varphi(u)$.
  \end{proof}

The next lemma states that if an agent receives an interior
allocation at some utility profile, then the rule is unresponsive to
non-ordinal changes at that 
utility  profile.
\begin{lem}\label{lem: ordinality if an interior}
    Suppose {$\varphi$   is       {efficient},  {strategy-proof},
    {non-bossy}, and  {\continuous}.} Let
  {$u, u'\in \mathcal U^N$ such that $P(u') = P(u)$.}
  {  If, for some $i\in N$,  $\nsupp{\varphi_{i}(u)} = 3$, then $\varphi(u') = \varphi(u)$.}
\end{lem}
\begin{proof}
Let $\pi = \varphi(u)$. Label the objects so that $u_{i}(a) > u_{i}(b) >
u_{i}(c)$.
Since $\nsupp{\pi_i} = 3$, for each $j\in N, \nsupp{\pi_j} \geq 2$.

For each $j\in N\setminus \{i\}$,
if $\pi_{ja} > 0$, then efficiency dictates that $u_{j}(a) > u_{j}(b)$
(otherwise $i$ would give $j$ a small amount of $b$ in exchange for
an equal amount of $a$ and both would be better off) and $u_{j}(a) >
u_{j}(c)$ (to preclude a similar trade of $c$ for $a$). Similarly, if
$\pi_{jb} > 0$ then efficiency dictates that $u_{j}(b) > u_{j}(c)$ (to
preclude a trade of $b$ for $c$). Thus, if there is $j$ such that
$\pi_{ja},\pi_{jb} > 0$, then $u_{j}(a) > u_{j}(b) > u_{j}(c)$. However,
since $\pi_{ib} > 0$, this
contradicts \cref{lem: same oder two agents}. Thus, for each $j\in
N\setminus\{i\}$,
$P(u_j)\neq P(u_i)$ and either $\pi_{ja} = 0$ or $\pi_{jb} = 0$ .

Since $\pi_{ia} <1$ there is $j$ such that $\pi_j\in (a,c)$. Since
$P(u_j) \neq P(u_i)$ and since $u_{j}(a) > u_{j}(b), u_{j}(c)$, we
conclude that $u_{j}(a) > u_{j}(c) >u_{j}(b)$.

By \cref{lem: lottery-ordinal utility}, we may suppose that $u_j\in
\overline{\mathcal U}$. Then, by \continuity, there is some  $u_j^j\in
\overline{\mathcal U}$ close to $u_j$ such that   $P(u_j^j) = P(u_j)$
and $\mu(u_j^j) \neq \mu(u_j)$ yet $\nsupp{\varphi_i(u_j^j, u_{-j})} =
3$. Let  $\pi^j = \varphi(u_j^j,
 u_{-j})$. By the reasoning above, $\pi_j^j\in (a,c)$. By
 strategy-proofness,  $\pi_j^j = \pi_j$ (otherwise, if $\pi^j_{ja} >
 \pi_{ja}$, $j$ would
 misreport $u_j^j$ at true utility $u_j$ and vice versa). Thus, by
 non-bossiness, $\varphi(u_j^j,u_{-j}) = \varphi(u)$. In other words,
 $\varphi$  responds only to $P(u_j)$ and not $\mu(u_j)$ at
 $u_{-j}$, so we may assume, without loss of generality, any
 $\mu(u_j)$.

There is $k\in N\setminus\{i,j\}$ such that
$\pi_{kb} > 0 $ since $\pi_{ib} < 1$ and $\pi_{jb}=0$. As observed
above, $u_{k}(b) > u_{k}(c)$. Also as observed above, $\pi_{ka} =
0$. Thus, $\pi_k\in (b,c)$. Since $P(u_k) \neq P(u_i)$, $u_{k}(b) >
u_{k}(a)$. By similar reasoning, we may also assume without loss of
generality any $\mu(u_k)$.

Consider $u_j'\in \mathcal U$ such that $P(u_j') = P(u_i)$. Let $\pi'
= \varphi(u_j', u_{-j})$. By \cref{lem: same oder two agents}, either
$\pi_{ib}' + \pi_{jb}' =0$ or $\pi_i',\pi_j' \in [a,b]\cup[b,c]$. If $\pi_{ib}' + \pi_{jb}'=0$,
then $\pi_j'\in [a,c]$. Then, by strategy-proofness, $\pi_j' = \pi_j$
(otherwise, if $\pi_{ja}' > \pi_{ja}$, $j$ would report $u_j'$ at true
preference $u_j$ and vice versa). By non-bossiness, then $\pi' =
\pi$, contradicting $\pi_{ib}' + \pi_{jb}'=0$. Thus, $\pi_i',
\pi_j'\in [a,b]\cup[b,c]$.

Note that by \cref{lem: same oder two agents}, $\varphi$ is invariant
to $\mu(u_j')$ at $u_{-j}$ as long as $u_{j}'(a) > u_{j}'(b) >
u_{j}'(c)$. Thus, we may assume  without loss of generality any $\mu(u_j')$.
If $\pi_{ja}' < \pi_{ja}$ then, for sufficiently low
$\mu(u_j')$ (that is, as if $u_{j}'(b)$ is close enough to $u_{j}'(c)$
given $u_{j}'(a)$) $j$ would report $u_j$ as true utility
$u_j'$. Conversely, if $\pi_{ja}' > \pi_{ja}$ then, for sufficiently low
$\mu(u_j)$ (that is, as if $u_{j}(c)$ is close enough to $u_{j}(b)$
given $u_{j}(a)$) $j$ would report $u_j'$ as true utility
$u_j$. Thus, $\pi_{ja}' = \pi_{ja}$.

Now, consider $\tilde u_i\in \overline{ \mathcal U}$ such that $P(\tilde u_i) =
P(u_i)$ and $|\mu(\tilde u_i) - \mu(u_i)| < \varepsilon$ for some
$\varepsilon > 0$. Let $\tilde \pi = \varphi(\tilde u_i, u_{-i})$.
By \continuity of $\varphi$, for small enough $\varepsilon$,
$\nsupp{\tilde \pi_i} = 3$. We now repeat the argument above starting
with the profile  $(\tilde u_i, u_{-i})$ rather than $u$ to conclude
that at $\tilde \pi$,   one agent in $N\setminus \{i\}$ receives an
assignment in $(a,c)$ and the other in $(b,c)$. By \continuity, the
former is $j$ and the latter is $k$. As above, we also conclude that
$\tilde \pi_{ka} = 0$.

Now let $\tilde \pi' = \varphi(\tilde u_i, u_j', u_k)$. Repeating the
argument above, $\tilde \pi_{ja}' = \tilde \pi_{ja}$. By \cref{lem:
  same oder two agents}, $\tilde \pi' = \pi'$, so $\pi_{ja} =
\pi'_{ja} = \tilde \pi_{ja}$. Since $\pi_{j}, \tilde \pi_{j}\in (a,c)$,
this implies $\pi_j = \tilde \pi_j$.
Since $\tilde \pi_{ka} = \pi_{ka} = 0$, this means $\tilde \pi_{ia} =
\pi_{ia}$. However, unless $\tilde \pi_i = \pi_i$, $i$ would have an
incentive to either misreport $u_i$ at the true utility $\tilde u_i$
(if $\tilde \pi_{ib} < \pi_{ib}$) or vice versa (if $\tilde \pi_{ib} >
\pi_{ib}$). Since $\tilde \pi_i = \pi_i$, and $\tilde \pi_j = \pi_j$,
we conclude that $\tilde \pi_k = \pi_k$. Thus, $\varphi$ does not
respond to $\mu(u_i)$ either.
\end{proof}

Our last lemma states that if an agent receives two objects with
positive probability, then the rule is unresponsive to non-ordinal
changes in that agent's utility.
\begin{lem}\label{lem: indiv ordinality if support leq two}
Suppose {$\varphi$   is       {efficient}, {strategy-proof},
      {non-bossy}, and {\continuous}. } Let
  {$u\in \mathcal U^N$ and $i\in N$ be such that $\nsupp{\varphi_i(u)}
    \leq 2$. If $u'_i\in \mathcal U$ is such that $P(u_i') = P(u_i)$,
    then $\varphi(u_i', u_{-i}) = \varphi(u)$.
    }
\end{lem}
\begin{proof}

  By \cref{lem: lottery-ordinal utility} we may assume that $u_i,
  u_i'\in\overline {\mathcal   U}$. Let $\pi = \varphi(u)$ and $\pi' =
  \varphi(u_i, u_{-i})$. By
  non-bossiness, if $\pi \neq \pi'$ then $\pi_i \neq \pi_i'$.
  Label the objects in $A$ so that $u_i(a) > u_i(b) > u_i(c)$.  If
  $\pi_{ia} = 1$ or $\pi_{ic} = 1$, by \cref{lem: top or bottom} we
  are done.  So, suppose that $\pi_i\in (a,b]\cup [b,c)\cup (a,c)$.

  For each
  $\alpha\in [0,1]$, let
  \[
    u_i^\alpha = \alpha u_i' + (1-\alpha)u_i\in   \overline {\mathcal
      U}\hspace{0.5in}\text{ and
    }\hspace{0.5in}\pi^\alpha =
    \varphi(u_i^\alpha, u_{-i}).
  \]
  By \cref{lem: ordinality if an interior}, for each $\alpha\in
  [0,1]$, $\nsupp{\pi^\alpha_i} < 3$. If $\pi_i = \pi_i^0 \in (a,b]\cup[b,c)$
  then for each $\alpha\in(0,1]$, $\pi^\alpha_i = \pi^0_i$. Otherwise,
  by \continuity,
  there is some $\underline \alpha$ such that $\pi_i^{\underline
    \alpha}\neq \pi_i^0$ but $\pi_i^{\underline \alpha} \in
  (a,b]\cup[b,c)$, contradicting strategy-proofness since $i$ either
  reports $u^{\underline \alpha}_i$ when their true utility is $u^0_i$
  or the other way around. Thus, since $\pi'_j = \pi^1_j  \neq
  \pi_i^0=\pi_i$, it must be that $\pi_i^0 \in (a,c)$. Again,
  \continuity and strategy-proofness imply then that for each $\alpha
  \in (0,1]$, $\pi_i^\alpha = \pi_i^0$, contradicting $\pi_i^1 \neq \pi_i^0$.
\end{proof}

We now wrap up the proof of \cref{ordinality theorem}.
\begin{proof}[Proof of \cref{ordinality theorem}]
  Take any pair $u, u'\in \mathcal U^N$ such that $P(u) = P(u')$.

  If there is
  an agent   $i\in N$ such that
  $\nsupp{\varphi_i(u)} = 3$, we conclude that $\varphi(u') =
  \varphi(u)$ by \cref{lem: ordinality if an interior}.
Otherwise,
  we conclude that $\varphi(u') =
  \varphi(u)$ by repeatedly applying \cref{lem: indiv ordinality if
    support leq two}  and non-bossiness.
\end{proof}

\section{Beyond Expected Utility}
\label{sec:nonEU}

\cref{ordinality theorem} tells us that a \continuous, efficient,
strategy-proof, and non-bossy rule ignores any non-ordinal information
contained in an agent's Bernoulli utility function. In this section, we
extend our result beyond expected utility. Specifically, we ask: Over
a larger domain of preferences, is there any additional information
that the rule can respond to?

We will consider  any domain $\mathcal V$ of utilities that satisfies
three conditions. The first two are:
\begin{enumerate}
\item\label{incl-eu} \textbf{Inclusion of expected utility:}  $\mathcal V \supseteq \mathcal
  U$.
\item\label{noties} \textbf{No ties on degenerate lotteries:} For each
  $i\in N$, each
  $u_i\in   \mathcal V$ and each distinct pair $a,b\in A, u_i(a) \neq u_i(b)$.
\end{enumerate}
Before we state the third condition, we introduce some notation. By
\ref{noties}. above,  for each $i\in N$ and $u_i\in\mathcal U$, the ordinal 
preference $P_i\in\mathcal P$ consistent with $u_i$ is well defined. For
each distinct pair of lotteries 
$\pi_i$ and $\pi'_i\in \Delta(A)$,  we say
$\bm{\pi_i}$ (first order) {\bf  stochastically dominates} $\bm{\pi'_i}$ {\bf
  at} $\bm{P_i}$ if for each $a\in A$, $\sum_{b P a}\pi_{ib}\ge
\sum_{b\mathrel  P a}\pi'_{ib}$. In this case, if $u_i\in \mathcal
V_i$ is consistent with $P_i$, we say that $\pi_i$ stochastically dominates $\pi'_i$ at
$u_I$ and denote it by $\pi_i \pisd \pi'_i$. If either $\pi_i \pisd
\pi'_i$ or $\pi_i =\pi'_i$, we write $\pi_i \risd \pi'_i$

Our third assumption on $\mathcal U$ is as  follows:
\begin{enumerate}
\setcounter{enumi}{2}
\item\label{sd}\textbf{Completion of  stochastic dominance:} For
  each $i\in N$, each $u_i\in\mathcal V$, and
  each pair of distinct lotteries $\pi_i, \pi'_i\in \Delta(A)$, if $\pi_i$
   stochastically dominates $\pi'_i$ at
  $u_i$, then $u_i(\pi) > u_i(\pi')$.
\end{enumerate}
Since every $u_i \in \mathcal U$ completes  stochastic
dominance and has no ties over degenerate lotteries, our assumptions
are compatible.

A rule on a domain $\mathcal V$ associates every profile of utilities
in $\mathcal V^N$ with an allocation in $\Pi$. The definitions of
efficiency, strategy-proofness, non-bossiness, and ordinality are
exactly as in 
\cref{sec:eff-axiom,sec:main-result} except that we replace $\mathcal
U$ with 
$\mathcal V$. Since 
$\overline{\mathcal U}$ is a subset of $\mathcal V$, the definition of
\continuity is unchanged.

\begin{thm}\label{ext ordinality theorem}
  Suppose that $|N| = |A| = 3$ and that  $\mathcal V$ satisfies
  \ref{incl-eu}, \ref{noties}, and \ref{sd} above.

  Every \continuous, efficient, strategy-proof,
  and non-bossy rule is ordinal.
\end{thm}
We prove a  lemma before proving 
\cref{ext ordinality theorem}. Though stochastic dominance comparisons
are 
incomplete, this lemma says that for every pair of uncomparable
lotteries, there are Bernoulli utilities consistent with
the same ordering that rationalize a preferences for one over the
other.

\begin{lem}\label{lem: uncomparable lotteries}
  Let $i\in N$ and $u_i\in\mathcal V$. Suppose that  $\pi_i^1,
  \pi_i^2 \in \Pi$ are such that it is not the case that
  $\pi_i^2 \risd \pi^1_i$. Then there is $u_i^1\in \mathcal U$
  such that $P(u_i^1) = P(u_i)$ and  $u_i^1(\pi_i^1)
  >u_i^1(\pi_i^2)$.
\end{lem}
\begin{proof}
  Since it is not the case that $\pi^2_i \risd \pi^1_i$,
there is $a\in A$ such that $\sum_{b: u_{i}(b) > u_{i}(a)}\pi^1_{ib}
-  \sum_{b: u_{i}(b) > u_{i}(a)}\pi^2_{ib} > \varepsilon$ for some
$\varepsilon > 0$.

Let $B =\{b\in A: u_{i}(b) > u_{i}(a)\}$, $\mu =
\max_{b\in B} u_{i}(a)$, and $\delta, \delta' > 0$. Define  $ u_i^1 \in
\mathcal U$ so that
\[
  u^1_{i}(b) = \left\{
    \begin{array}{ll}
      1- \delta\left(\frac{\mu - u_{i}(b)}{\mu - u_{i}(a)}\right)
      &
        \text{if }b\in B\\
      \delta'u_{i}(b) &\text{if }b\notin B.
    \end{array}
    \right.
  \]
For  $\delta, \delta'$ small enough, $P(u_i^1) = P(u_i)$
  and $u^1_i(\pi_i^1) > u_i^1(\pi_i^2)$.
\end{proof}

\begin{proof}[Proof of \cref{ext ordinality theorem}]
  Suppose that $\varphi$ is \continuous, efficient, strategy-proof, and
  non-bossy. Let $u, u'\in \mathcal V^N$ be such that $P(u) =
  P(u')$. We need to  show that $\varphi(u) = \varphi(u')$.

  First consider the case where  $u,  u'\in \mathcal U^N$. We conclude
  by  \cref{ordinality
    theorem} that  $\varphi(u) = \varphi(u')$.

  Second, consider  $u\in \mathcal U^N$ and $u'\notin
  \mathcal 
  U^N$. Let $\pi^1 = \varphi(u'_1, u_{-i})$. If $\pi_1^1 \neq \pi_1$,
  then by strategy-proofness, it
  cannot be that   $\pi^1_1 \risd \pi_1$. Therefore, by \cref{lem:
    uncomparable lotteries} there   is $u^1_1\in \mathcal U$  such
  that $P(u_1^1) = P(u_1') = P(u_1)$ and $u_1^1(\pi_1^1) >
  u_1^1(\pi_1)$. However, $(u^1_1, u_{-1}) \in \mathcal U^N$, so the
  first case  implies that   $\varphi(u^1_1, 
   u_{-1}) = 
  \pi$. However, since $u^1_1(\varphi_1(u_1',
  u_{-1})) = u^1_1(\pi^1_1)   > u^1_1(\pi_1) = u_1^1(\varphi_1(u^1_1,
  u_{-1})$, this contradicts strategy-proofness. Thus, $\pi_1^1 =
  \pi_1$ and by non-bossiness, $\pi^1 = \pi$. Repeating the argument
  for the remaining agents one at a time, we conclude that
  $\varphi(u') = \varphi(u)$.

  Finally, consider $u,u'\in \mathcal U^N\setminus \mathcal  U^N$. Let
  $\overline u\in \mathcal U^N$ be such that $P(\overline u) = P(u) =
  P(u')$. By the above argument, $\varphi(u) = \varphi(\overline u) =
  \varphi(u')$.  
\end{proof}

\bibliography{refs.bib}

\end{document}